\documentclass[twoside,12pt]{article}

\usepackage{epsfig}

\def\NPA{{\em Nucl. Phys.} {\bf A}}

\def\PLB{{\em Phys. Lett.} {\bf B}}

\def\PRL{{\em Phys. Rev. Lett.} }

\def\PRC{{\em Phys. Rev.} {\bf C}}

\newcommand{\be}{\begin{equation}}
\newcommand{\ee}{\end{equation}}
\newcommand{\bea}{\begin{eqnarray}}
\newcommand{\eea}{\end{eqnarray}}

\begin{document}
\title{ \vspace{1cm} In-medium Properties of Hadrons -- Observables II}
\author{L.\ Alvarez-Ruso, T.\ Falter, U.\ Mosel, P.\ Muehlich
\\
Institut fuer Theoretische Physik, Universitaet Giessen\\
D-35392 Giessen, Germany}
\maketitle
\begin{abstract}
In this review we discuss the observable consequences of in-medium
changes of hadronic properties in reactions with elementary
probes, and in particular photons, on nuclei. After an outline of
the theoretical method used we focus on a discussion of actual
observables in photonuclear reactions; we discuss in detail $2\pi
$- and vector-meson production. We show that the $2\pi^0$
photoproduction data can be well described by final state
interactions of the pions produced whereas the semi-charged
$\pi^0\pi^\pm$ channel exhibits a major discrepancy with theory.
For $\omega$ production on nuclei in the TAPS/CB@ELSA experiment
we analyse the $\pi^0\gamma$ decay channel, and illustrate the
strength of the method by simulating experimental acceptance
problems. Completely free of final state interactions is dilepton
production in the few GeV range. We show that the sensitivity of
this decay channel to changes of hadronic properties in medium in
photonuclear reactions on nuclei is as large as in
ultrarelativistic heavy ion collisions and make predictions for
the on-going G7 experiment at JLAB. Finally we discuss that hadron
production in nuclei at 10 -- 20 GeV photon energies can give
important information on the hadronization process, and in
particular on the time-scales involved. We show here detailed
calculations for the low-energy (12 GeV) run at HERMES and
predictions for planned experiments at JLAB.

\end{abstract}


\section{Introduction}

Studies of in-medium properties of hadrons are driven by a number
of - partly connected - motivations. First, there is the
expectation that changes of hadronic properties in medium can be
precursor phenomena to changes in the phase structure of nuclear
matter. Here the transition to the chirally symmetric phase, that
exhibits manifestly the symmetries of the underlying theory of
strong interactions, i.e. QCD, is of particular interest. Present
day's ultrarelativic heavy-ion collisions explore this phase
boundary in the limit of high temperatures ($T \approx$ 170 MeV)
and low densities. The other limit (low temperatures and high
densities) is harder to reach, although the older AGS heavy-ion
experiments and the planned CBM experiment at the new FAIR
facility \cite{FAIR} may yield insight into this area. However,
even in these experiments the temperatures reached are still
sizeable ($T \approx $ 120 MeV). At very low temperatures the only
feasible method seems to be the exploration of the hadronic
structure inside ordinary nuclei. Here the temperature is $T=0$
and the density at most equals the equilibrium density of nuclear
matter, $\rho_0$. It is thus of great interest to explore if such
low densities can already give precursor signals for chiral
symmetry restoration. The second motivation for the study of
hadronic in-medium properties is provided by our interest in
understanding the structure of large dense systems, such as the
interior of stars. This structure obviously depends on the
composition of stellar matter and its interactions.

In \cite{I} we have discussed the relevant questions and
theoretical studies of in-medium properties in some detail. Such
calculations necessarily rely on a number of simplifying
assumptions, foremost being that of an infinite medium at rest in
which the hadron under study is embedded. In actual experiments
these hadrons are observed through their decay products and these
have to travel through the surrounding nuclear matter to the
detectors. Except for the case of electromagnetic signals
(photons, dileptons) this is connected with often sizeable final
state interactions (FSI) that have to be treated as realistic as
possible. For a long period the Glauber approximation which allows
only for absorptive processes along a straight-line path has been
the method of choice in theories of photonuclear reactions on
nuclei. This may be sufficient if one is only interested in total
yields. However, it is clearly insufficient when one aims at, for
example, reconstructing the spectral function of a hadron inside
matter through its decay products. Rescattering and sidefeeding
through coupled channel effects can affect the final result so
that a realistic description of such effects is absolutely
mandatory \cite{FalterShad}.

In the present paper we present the method used for the full event
simulation in some more details while being quite short on the
theoretical calculations of in-medium properties. For the latter
we refer to \cite{I}.

That hadrons can change their properties and couplings in the
nuclear medium has been well known since the days of the
Delta-hole model that dealt with the changes of the properties of
the pion and Delta-resonance inside nuclei \cite{Ericsson-Weise}.
Due to the predominant $p$-wave interaction of pions with nucleons
one observes a lowering of the pion branch for small, but finite
pion momenta, which increases with the nucleon-density. More
recently, experiments at the FRS at GSI have shown that also the
pion rest mass in medium differs from its value in vacuum
\cite{Kienle}.

In addition, vector mesons have found an increasing interest over
the last few years. These mesons are the first excitations of the
QCD vacuum that are not protected by chiral symmetry, as the pions
as Goldstone bosons are, and should therefore be more directly
related to the chiral condensates. Indeed, the original work by
Hatsuda and Lee \cite{Hatsuda} based on QCD sum rules predicted a
significant lowering of the vector meson masses with increasing
density, the effect being as large as 30\% already at nuclear
matter equilibrium density. As discussed in more detail in
\cite{I} and in particular in \cite{Leupold,LeupoldMosel} this
original prediction strongly depended on simplifying assumptions
for the spectral function of the particles involved. When more
realistic spectral shapes are used the QCD sum rule gives only
certain restrictions on mass and width of the particles involved,
but does not fix the latter; for that hadronic models are needed.
In particular, for the $\rho$ meson it turned out that the
broadening is more dominant than a mass-shift \cite{Postneu}. This
scenario is in line with observations by the CERES experiment
\cite{CERES} that has found a considerable excess of dileptons in
an invariant mass range from $\approx 300$ MeV to $\approx 700$
MeV as compared to expectations based on the assumption of freely
radiating mesons. This result has found an explanation in terms of
a shift of the $\rho$ meson spectral function down to lower
masses, as expected from theory (see, e.g., \cite{Postneu,
Peters,Post,Wambach}). However, while quite different model
calculations tend to explain the data, though often with some
model assumptions \cite{Cassingdil,RappWam,Rapp,Koch,Renk} their
theoretical input is sufficiently different as to make the inverse
conclusion that the data prove one or another of these scenarios
impossible. The more recent experimental results on a change of
the rho-meson properties in-medium obtained in an
ultrarelativistic $Au + Au$ collision by the STAR collaboration at
RHIC \cite{STAR} show a downward shift of the $\rho$-meson pole
mass by about 70 MeV. Since also the $p + p$ data obtained in the
same experiment show a similar, though slightly less pronounced
(-40 MeV), shift of the $\rho$-meson mass, phase-space distortions
of the $\rho$-meson spectral shape may be at least partly
responsible for the observed mass shift. In the heavy-ion
experiment then a number of additional effects, mainly by
dynamical interactions with surrounding matter, may contributed,
but are hard to separate from the more mundane phase space
effects.

One of the authors has, therefore, already some years ago proposed
to look for the theoretically predicted changes of vector meson
properties inside the nuclear medium in reactions on normal nuclei
with more microscopic probes \cite{Hirschegg97,Erice98,Hirschegg}.
Of course, the average nuclear density felt by the vector mesons in such
experiments lies much below the equilibrium density of nuclear
matter, $\rho_0$, so that naively any density-dependent effects
are expected to be much smaller than in heavy-ion reactions.

On the other hand, there is a big advantage to these experiments:
they proceed with the spectator matter being close to its
equilibrium state. This is essential because all theoretical
predictions of in-medium properties of hadrons are based on an
model in which the hadron (vector meson) under investigation is
embedded in nuclear matter in equilibrium and with infinite
extension. However, a relativistic heavy-ion reaction proceeds --
at least initially -- far from equilibrium. Even if equilibrium is
reached in a heavy-ion collision this state changes by cooling
through expansion and particle emission and any observed signal is
built up by integrating over the emissions from all these
different stages of the reaction.

Another in-medium effect arises when particles are produced by
high-energy projectiles inside a nuclear medium. A major experimental
effort at RHIC experiments has gone into the observation of jets in
ultra-relativistic heavy-ion collisions and the determination of
their interaction with the surrounding quark or hadronic matter \cite{Jet}.
A complementary process is given by the latest HERMES results at HERA
for high-energy electroproduction of hadrons off nuclei \cite{Hermes}.
Again, the advantage of the latter experiment is that the nuclear matter
with which the interactions happen is at rest and in equilibrium.

In this lecture note we summarize results that we have obtained in
studies of observable consequences of in-medium changes of
hadronic spectral functions as well as hadron formation in
reactions of elementary probes with nuclei. We demonstrate that
the expected in-medium sensitivity in such reactions is as high as
that in relativistic heavy-ion collisions and that in particular
photonuclear reactions present an independent, cleaner testing
ground for assumptions made in analyzing heavy-ion reactions.

\section{Theory}

\subsection{QCD Sum Rules and Hadronic Models}
A review of the underlying theory can be found in \cite{I}. Here
we would just like to mention that the connection between chiral
condensates on one hand and hadronic observables on the other
cannot simply be inferred from looking at the dependence of the
condensate on density and temperature. Instead, the connection is
much more indirect; only an integral over the spectral function
can be linked via the QCD sum rule to the condensate behavior in
medium. In an abbreviated form this connection is given by
\begin{eqnarray}
R^{{\rm OPE}}(Q^2) & = & {\tilde c_1 \over Q^2} + \tilde c_2 -{Q^2
\over \pi} \int\limits_0^\infty \!\! ds \, {{\rm Im}R^{{\rm
HAD}}(s) \over (s+Q^2)s}
  \label{eq:opehadr}
\end{eqnarray}
with $Q^2:= -q^2 \gg 0$ and some subtraction constants $\tilde
c_i$. Here $R^{\rm OPE}$ represents a Wilson's operator expansion
of the current-current correlator in terms of quark and gluon
degrees of freedom in the space-like region. It is an expansion in
terms of powers of $1/Q^2$ and contains the condensates as
expansion parameters. On the other hand, $R^{{\rm HAD}}(s)$ in
(\ref{eq:opehadr}) is the same object for time-like momenta,
represented by a parametrization in terms of hadronic variables.
The dispersion integral connects time- and space-like momenta. Eq.
(\ref{eq:opehadr}) connects the hadronic with the quark world. It
allows to determine parameters in a hadronic parametrization of
$R^{{\rm HAD}}(s)$ by comparing the lhs of this equation with its
rhs. Invoking vector meson dominance it is easy to see that for
vector mesons Im$R^{{\rm HAD}}(s)$ in (\ref{eq:opehadr}) is just
the spectral function of the meson under study.

The operator product expansion of $R^{\rm OPE}$ on the lhs
involves quark- and gluon condensates \cite{Leupold,LeupoldMosel};
of these only the two-quark condensates are reasonably well known,
whereas our knowledge about already the four-quark condensates is
rather sketchy.

Using the measured, known vacuum spectral function for $R^{\rm
HAD}$ allows one to obtain information about these condensates
appearing on the lhs of (\ref{eq:opehadr}). On the other hand,
modelling the density-dependence of the condensates yields
information on the change of the hadronic spectral function when
the hadron is embedded in nuclear matter. Since the spectral
function appears under an integral the information obtained is not
direct. Therefore, as Leupold et al.\ have shown
\cite{Leupold,LeupoldMosel}, the QCDSR provides important
constraints for the hadronic spectral functions in medium, but it
does not fix them. Recently Kaempfer et al have turned this
argument around by pointing out that measuring an in-medium
spectral function of the $\omega$ meson could help to determine
the density dependence of the higher-order condensates
\cite{Kaempfer}.

Thus models are needed for the hadronic interactions. The
quantitatively reliable ones can at present be based only on
'classical' hadrons and their interactions. Indeed, in lowest
order in the density the mass and width of an interacting hadron
in nuclear matter at zero temperature and vector density $\rho_v$
are given by (for a meson, for example)
\begin{eqnarray}     \label{trho}
{m^*}^2 & =  & m^2 - 4 \pi \mathcal{R}e f_{m N}(q_0,\theta = 0)\, \rho_v
\nonumber \\
m^* \Gamma^* & = & m \Gamma^0 -  4 \pi \mathcal{I}m f_{mN}(q_0,\theta = 0)\,
\rho_v ~.
\end{eqnarray}
Here $f_{mN}(q_0,\theta = 0)$ is the forward scattering amplitude
for a meson with energy $q_0$ on a nucleon. The width $\Gamma^0$
denotes the free decay width of the particle. For the imaginary
part this is nothing other than the classical relation $\Gamma^* -
\Gamma^0 = v \sigma \rho_v$ for the collision width, where
$\sigma$ is the total cross section. This can easily be seen by
using the optical theorem.

Unfortunately it is not a-priori known up to which densities the
low-density expansion (\ref{trho}) is useful. Post et al.
\cite{Postneu} have recently investigated this question in a
coupled-channel calculation of selfenergies. Their analysis
comprises pions, $\eta$-mesons and $\rho$-mesons as well as all
baryon resonances with a sizeable coupling to any of these mesons.
The authors of \cite{Postneu} find that already for densities less
than $0.5 \rho_0$ the linear scaling of the selfenergies inherent
in (\ref{trho}) is badly violated for the $\rho$ and the $\pi$
mesons, whereas it is a reasonable approximation for the $\eta$
meson. Reasons for this deviation from linearity are Fermi-motion,
Pauli-blocking, selfconsistency and short-range correlations. For
different mesons different sources of the discrepancy prevail: for
the $\rho$ and $\eta$ mesons the iterations act against the
low-density theorem by inducing a strong broadening for the
$D_{13}(1520)$ and a slightly repulsive mass shift for the
$S_{11}(1535)$ nucleon resonances to which the $\rho$ and the
$\eta$ meson, respectively, couple. The investigation of in-medium
properties of mesons, for example, thus involves at the same time
the study of in-medium properties of nucleon resonances and is
thus a coupled-channel problem.

\subsection{Coupled Channel Treatment of Incoherent Particle
Production}\label{sec:CoupledChannel}

Very high nuclear densities ($2 - 8\rho_0$) and temperatures $T$
up to or even higher than $\approx$ 170 MeV can be reached with
present day's accelerators in heavy-ion collisions. Thus, any
density-dependent effect gets magnified in such collisions.
However, the observed signal always represents a time-integral
over various quite different stages of the collision --
non-equilibrium and equilibrium, the latter at various densities
and temperatures. The observables thus have to be modelled in a
dynamic theory. In contrast, the theoretical input is always
calculated under the simplifying assumption of a hadron in
stationary nuclear matter in equilibrium and at fixed density. The
results of such calculations are then used in dynamical
simulations of various degrees of sophistication most of which
invoke a quasi-stationary approximation. In order to avoid these
intrinsic difficulties we have looked for possible effects in
reactions that proceed closer to equilibrium, i.e. reactions of
elementary probes such as protons, pions, and photons on nuclei.
The densities probed in such reactions are always $\le \rho_0$,
with most of the nucleons actually being at about $0.5 \rho_0$. On
the other hand, the target is stationary and the reaction proceeds
much closer to (cold) equilibrium than in a relativistic heavy-ion
collision. If any observable effects of in-medium changes of
hadronic properties survive, even though the densities probed are
always $\le \rho_0$, then the study of hadronic in-medium
properties in reactions with elementary probes on nuclei provides
an essential baseline for in-medium effects in hot nuclear matter
probed in ultra-relativistic heavy-ion collisions.

With the aim of exploring this possibility we have over the last
few years undertaken a number of calculations for proton-
\cite{Bratprot}, pion- \cite{Weidmann,Effepi} and photon-
\cite{Effephot} induced reactions. All of them have one feature in
common: they treat the final state incoherently in a coupled
channel transport calculation that allows for elastic and
inelastic scattering of, particle production by and absorption of
the produced hadrons. A new feature of these
calculations is that hadrons with their correct spectral functions
can actually be produced and transported consistently. This is
quite an advantage over earlier treatments \cite{Brat-Cass, Fuchs}
in which the mesons were always produced and transported with
their pole mass and their spectral function was later on folded in
only for their decay. The method is summarized in the following
section, more details can be found in \cite{Effephot}.

We separate the photonuclear reaction into three steps. First, we
determine the amount of shadowing for the incoming photon; this
obviously depends on its momentum transfer $Q^2$. Second, the
primary particle is produced and third, the produced particles are
propagated through the nuclear medium until they leave the
nucleus.

\begin{figure}[h]
\begin{center}

\begin{minipage}[t]{8 cm}
\centerline{\epsfig{file=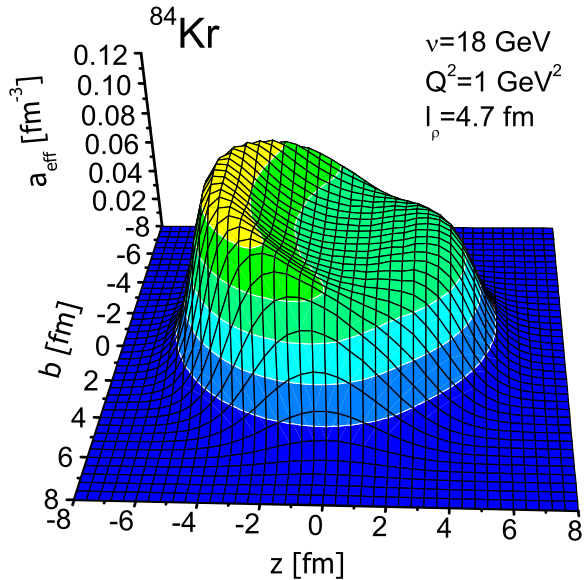,scale=1.2}}
\end{minipage}
\begin{minipage}[t]{16.5cm}
\caption{Profile function for shadowing. Shown is the probability
distribution for the interaction of an incoming photon (from left)
with given virtuality and energy with nucleons in a $^{208}Pb$
nucleus (from \cite{FalterShad}).}
\end{minipage}

\end{center}
\label{fig:shadow}
\end{figure}

\paragraph{Shadowing.} Photonuclear reactions show shadowing in the
entrance channel, for real photons from an energy of about 1 GeV
on upwards \cite{Bianchi}. This shadowing is due to a coherent
superposition of bare photon and vector meson components in the
incoming photon and is handled here by means of a Glauber multiple
scattering model \cite{FalterShad}. In this way we obtain for each
value of virtuality $Q^2$ and energy $\nu$ of the photon a spatial
distribution for the probability that the incoming photon reaches
a given point; for details see \cite{FalterShad,Effe,Falterinc}.
Fig.\ \ref{fig:shadow} gives an example of the spatial
distribution for a kinematical situation which is close to that of
the HERMES experiment.
The function shown in Fig. \ref{fig:shadow} gives the probability
distribution that the incoming photon hits the nucleon at that
point.

\paragraph{Initial Production.}
The initial particle production is handled differently depending
on the invariant mass $W = \sqrt{s}$ of the excited state of the
nucleon. If $W < 2$ GeV, we invoke a nucleon resonance model that
has been adjusted to nuclear data on resonance-driven particle
production \cite{Effephot}. If $W > 2$ GeV the particle yield is
calculated with standard codes developed for high energy nuclear
reactions, i.e.\ FRITIOF or PYTHIA; details are given in
\cite{Falterneu}. We have made efforts to ensure a smooth
transition of cross sections in the transition from resonance
physics to DIS.

\paragraph{Final State Interactions.}
The final state is described by a semiclassical coupled channel
transport model that had originally been developed for the
description of heavy-ion collisions and has since then been
applied to various more elementary reactions on nuclei with
protons, pions and photons in the entrance channel.

In this method the spectral phase space distributions of all
particles involved are propagated in time, from the initial first
contact of the photon with the nucleus all the way to the final
hadrons leaving the nuclear volume on their way to the detector.
The spectral phase space distributions
$F_h(\vec{x},\vec{p},\mu,t)$ give at each moment of time and for
each particle class $h$ the probability to find a particle of that
class with a (possibly off-shell) mass $\mu$ and momentum
$\vec{p}$ at position $\vec{x}$. Its time-development is determined
by the BUU equation
\be     \label{BUU}
(\frac{\partial}{\partial t} + \frac{\partial H_h}{\partial
\vec{p}} \frac{\partial}{\partial \vec{r}} - \frac{\partial
H_h}{\partial \vec{r}} \frac{\partial}{\partial \vec{p}})F_h=G_h
{\cal A}_h - L_h F_h.
\ee
Here $H_h$ gives the energy of the hadron $h$ that is being
transported; it contains the mass, the selfenergy (mean field) of
the particle and a term that drives an off-shell particle back to
its mass shell. The terms on the lhs of (\ref{BUU}) are the
so-called \emph{drift terms} since they describe the independent
transport of each hadron class $h$. The terms on the rhs of
(\ref{BUU}) are the \emph{collision terms}; they describe both
elastic and inelastic collisions between the hadrons. Here the
term \emph{inelastic collisions} includes those collisions that
either lead to particle production or particle absorption. The
former is described by the \emph{gain term} $G_h {\cal A}_h$ on
the rhs in (\ref{BUU}), the latter process (absorption) by the
\emph{loss term} $L_h F_h$. Note that the gain term is
proportional to the spectral function $\mathcal{A}$ of the
particle being produced, thus allowing for production of off-shell
particles. On the contrary, the loss term is proportional to the
spectral phase space distribution itself: the more particles there
are the more can be absorbed. The terms $G_h$ and $L_h$ on the rhs
give the actual strength of the gain and loss terms, respectively.
They have the form of Born-approximation collision integrals and
take the Pauli-principle into account. The free collision rates
themselves are taken from experiment or are calculated
\cite{Effephot}.

Eq.\ (\ref{BUU}) contains a selfconsistency problem. The collision
rates embedded in $G$ and $L$ determine the collisional broadening
of the particles involved and thus their spectral function
$\mathcal{A}$. The widths of the particles, resonances or mesons,
thus evolve in time away from their vacuum values. In addition,
broad particles can be produced off their peak mass and then
propagated. The extra 'potential' in $H$ already mentioned ensures
that all particles are being driven back to their mass-shell when
they leave the nucleus. The actual method used is described in
\cite{Effephot}. It is based on an analysis of the Kadanoff-Baym
equation that has led to practical schemes for the propagation of
off-shell particles \cite{Leupoldoffshell,JuchemCassing}´. The
possibility to transport off-shell particles represents a major
breakthrough in this field. For further details of the model see
Ref. \cite{Effephot} and \cite{Falterneu} and references therein.

\section{Particle Production on Nuclei -- Observables}

\section{\it $\eta$ Production} We first look at the prospects of
using reactions with hadronic final states and discuss the
photo-production of $\eta$-mesons on nuclei as a first example.
These mesons are unique in that they are sensitive to one
dominating resonance the $S_{11}(1535)$ so that one may hope to
learn something about the properties of this resonance inside
nuclei. Experiments for this reaction were performed both by the
TAPS collaboration \cite{TAPS,Krusche} and at KEK \cite{KEK}.

Estimates of the collisional broadening of the $S_{11}(1535)$
resonance have given a collision width of about 35 MeV at $\rho_0$
\cite{EffeHom}. The more recent, and more refined, selfconsistent
calculations of \cite{Postneu} give a very similar value for this
resonance. In addition, a dispersive calculation of the real part
of the selfenergy for the resonance at rest gives only an
insignificant shift of the resonance position. Thus any
momentum-dependence of the selfenergy as observed in
photon-nucleus data can directly be attributed to binding energy
effects \cite{Lehreta}. The results obtained in \cite{Lehreta}
indicate that the momentum dependence of the $N^*(1535)$ potential
has to be very similar to that of the nucleons.

\begin{figure}[h]

\begin{center}

\begin{minipage}[t]{16.5 cm}
\centerline{\epsfig{file=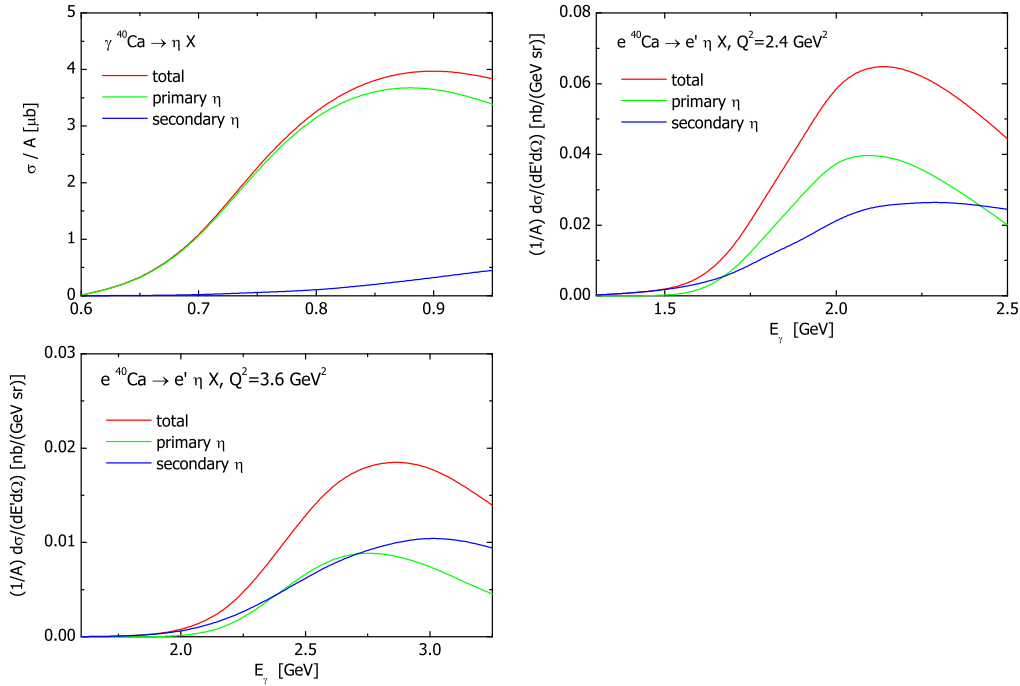,scale=1.2}}
\caption{Eta-electroproduction cross sections on $^{40}Ca$ for
three values of $Q^2$ given in the figures. The uppermost curve in
all three figures gives the total observed yield, the next curve
from top gives the contribution of the direct $\eta$ production
channel and the lowest curve shows the contribution of the
secondary process $\pi N \rightarrow \eta N$. For the highest
value of $Q^2=3.6$ GeV$^2$ the primary and secondary production
channels nearly coincide (from
\cite{Lehrelectro}).}\label{fig:electroeta}
\end{minipage}

\end{center}

\end{figure}

The particular advantage of our coupled-channels approach can be
seen in Fig.\ \ref{fig:electroeta} which shows results both for
photon- and electroproduction of $\eta$'s on nuclei; for the
latter process so far no data are available.
The calculations give the interesting result that for
photo-production a secondary reaction channel becomes important at
high energies and virtualities. In this channel first a pion is
produced which travels through the nucleus and through $\pi N
\rightarrow N^*(1535) \rightarrow N \eta$ produces the $\eta$ that
is finally observed in experiment. This channel becomes even
dominant for high $Q^2$ electroproduction. In both cases the
reason for the observed growing of the importance of secondary
production channels lies in the higher momentum transferred to the
initial pion \cite{Lehrelectro}.

\subsection{$2\pi$ Production}
If chiral symmetry is restored, the masses of the scalar
isoscalar $\sigma$ meson and that of the scalar isovector pion
should become degenerate. This implies that the spectral function
of the $\sigma$ should become softer and narrower with its
strength moving down to the $2\pi$ threshold. This leads to a threshold
enhancement in the $\pi\pi$ invariant mass spectrum due to suppression
of the phase-space for the $\sigma\rightarrow\pi\pi$ decay.

A first measurement of the two pion invariant mass spectrum has been
obtained by the CHAOS collaboration in pion induced reactions on
nuclei \cite{CHAOS}. The authors of \cite{CHAOS} claimed to indeed have
seen an accumulation of spectral strength near the $2\pi$ threshold for
heavy target nuclei in the $\pi^+\pi^-$ mass distribution. According to the
arguments presented in the introduction, photon induced reactions in nuclei
are much better suited to investigate double pion pion production at finite
baryon densities. A recent
experiment with the TAPS spectrometer at the tagged-photon facility MAMI-B in
Mainz indeed shows an even more pronounced accumulation of spectral strength of the two
pion mass spectrum for low invariant masses with increasing target
mass corresponding to increasing average densities probed
\cite{TAPSsigma}. This accumulation has been observed for the
$\pi^0\pi^0$ but not for the $\pi^\pm \pi^0$ final state.

These results have been explained by a model developed by Roca et al.
\cite{Oset2pi}.  In this model the $\sigma$ meson is generated dynamically
as a resonance in the $\pi\pi$ scattering amplitude. By dressing the pion
propagators in the medium by particle-hole loops, they found the observed
downward shift of the $\pi\pi$ mass spectrum to be consistent with a dropping of
the $\sigma$-pole in the $\pi\pi$ scattering amplitude, i.~e. a lowering of
the $\sigma$ meson mass.

\begin{figure}[h]
\begin{center}
\begin{minipage}[t]{10 cm}
\centerline{\epsfig{file=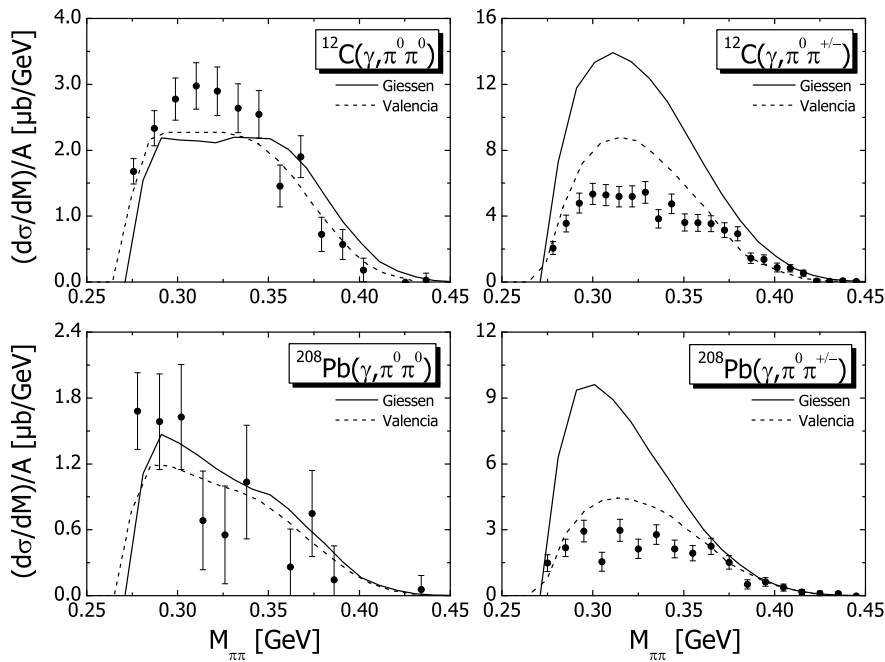,scale=0.8}}
\end{minipage}
\begin{minipage}[t]{16.5 cm}
\caption{Two pion invariant mass distributions for the
$\pi^0\pi^0$ (left) and $\pi^0\pi^\pm$ (right) photoproduction on
$^{12}C$ and $^{208}Pb$ (from \cite{Ruso}). The solid lines
represent our results whereas the dashed lines labelled
``Valencia'' depict the results of ref. \cite{Oset2pi}.}
 \label{fig:2pi}
\end{minipage}
\end{center}
\end{figure}

In analyzing the TAPS data it is absolutely essential to simulate
the final state interactions of the outgoing pions, correlated or
not, in an as realistic way as possible. We have, therefore, taken
the conservative approach of analyzing this reaction without any
changes of the $\sigma$ spectral function in the
outgoing channel. The result of this study \cite{Ruso} is shown in
Fig.~\ref{fig:2pi}.

Fig.~\ref{fig:2pi} shows on its left side that the observed
downward shift of the $2\pi^0$ mass spectrum can be well
reproduced by final state interactions on the independent pions
without any in-medium modification of the $\pi\pi$ interaction.
This shift can be attributed to a
slowing down of the pions due to quasi elastic collisions with the
nucleons in the surrounding nuclear medium. Also shown is the result of the
calculations of Roca et al. \cite{Oset2pi}. Both
calculations obviously agree with each other so that the final
observables are fairly independent of any $\pi\pi$ correlations.

On the right side of Fig.~\ref{fig:2pi} the results for the
semi-charged $\pi^0\pi^\pm$ channel are shown. Here our
theoretical result (solid line) overestimates the data by up to a
factor of 3. The result obtained in Ref.~\cite{Oset2pi} also lies
too high, but by a lesser amount; this has been explained in
\cite{Ruso} by the neglect of charge-transfer channels in the
calculations of Roca et al. \cite{Oset2pi}. The observed
discrepancy with experiment for this channel is astounding since
the method used normally describes data within a much narrower
error band. Thus understanding this discrepancy is absolutely
essential before a shift or non-shift in the mass distribution for
this channel can be ascertained.

\subsection{\it $\omega$ Production}

Many of the early studies of hadronic properties in medium
concentrated on the $\rho$-meson \cite{Peters,Klingl}, partly
because of its possible significance for an interpretation of the
CERES experiment. It is clear by now, however, that the dominant
effect on the in-medium properties of the $\rho$-meson is
collisional broadening that overshadows any possible mass shifts
\cite{Postneu} and is thus experimentally hard to observe. The
emphasis has, therefore, shifted to the $\omega$ meson. An
experiment measuring the $A(\gamma,\omega \rightarrow
\pi^0\gamma')X$ reaction is presently being analyzed by the
TAPS/Crystal Barrel collaborations at ELSA \cite{Messch}. The
varying theoretical predictions for the $\omega$ mass (640-765
MeV) \cite{Klingl} and width (up to 50 MeV) \cite{Weidmann,Friman}
in nuclear matter at rest encourage the use of such an exclusive
probe to learn about the $\omega$ spectral distribution in nuclei.
\begin{figure}[h]
\begin{center}
\begin{minipage}[t]{10 cm}
\centerline{\epsfig{file=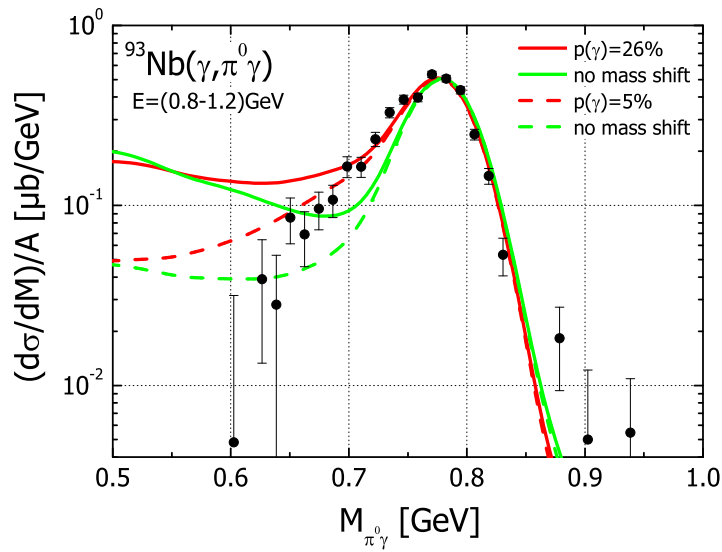,scale=1.0}}
\end{minipage}
\begin{minipage}[t]{16.5 cm}
\caption{Mass differential cross section for $\pi^0\gamma$
photoproduction off $^{93}$Nb. Shown are results both with and
without a mass-shift as explained in \cite{MuehlOm}. The quantity
$p$ gives the escape probability for one of the four photons in
the $2\pi^0$ channel. The two solid curves give results of
calculations with p = 26 \% with and without mass-shift, the two
dashed curves give the same for p = 5\% \label{omega}}
\end{minipage}
\end{center}
\end{figure}

Simulations have been performed at 1.2 GeV and 2.5 GeV photon
energy, which cover the accessible energies of the TAPS/Crystal
Barrel experiment. After reducing the combinatorial and
rescattering background by applying kinematic cuts on the outgoing
particles, we have obtained rather clear observable signals for an
assumed dropping of the $\omega$ mass inside nuclei
\cite{MuehlOm}. Therefore, in this case it should be possible to
disentangle the collisional broadening from a dropping mass, which
may be related to a change of the chiral condensate at finite
nuclear density \cite{BrownRho}.

Our calculations represent complete event simulations. It is,
therefore, possible to take experimental acceptance effects into
account. An example is shown in Fig.~\ref{omega} which shows the
effects of a possible misidentification of the $\omega$ meson.
This misidentification can come about through the $2 \pi^0
\rightarrow 4 \gamma$ channel if one of the four photons escapes
detection and the remaining three photons are identified as
stemming from the $\pi^0 \gamma \rightarrow 3 \gamma$ decay
channel of the $\omega$-meson.

Fig.\ \ref{omega} shows a good agreement between the preliminary
data of the TAPS/CB@ELSA collaboration \cite{Trnka} for a photon
escape probability of 5 \% and a mass shift $m_\omega = m_\omega^0
- 0.18 \,\rho/\rho_0$. In \cite{MuehlOm, I} we have also discussed
the momentum-dependence of the $\omega$-selfenergy in medium and
have pointed out that this could be accessible through
measurements which gate on different three-momenta of the $\omega$
decay products.

\subsection{\it Dilepton Production}

\begin{figure}[h]

\begin{center}

\begin{minipage}[t]{8 cm}
\centerline{\epsfig{file=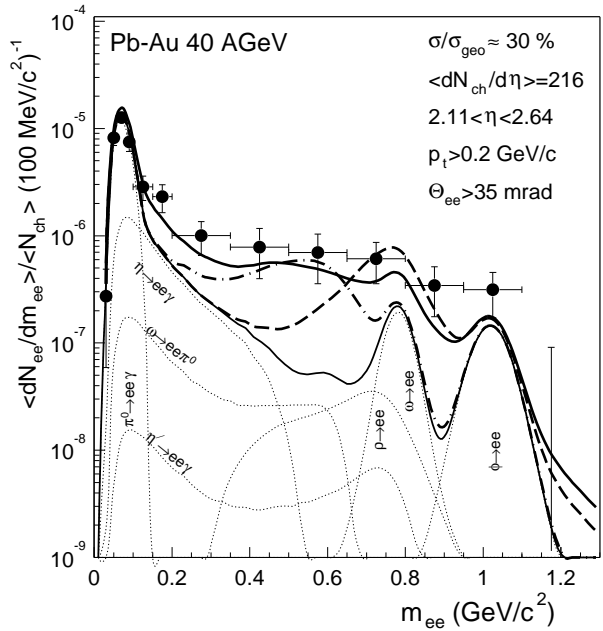,scale=0.8}}
\end{minipage}
\hspace{-0.5cm}
\begin{minipage}[t]{16.5 cm}
\caption{Invariant  dilepton mass spectrum
 obtained with the CERES experiment in Pb + Au collisions at 40
 AGeV (from \cite{CERES}). The thin curves give the contributions
 of individual hadronic sources to the total dilepton yield, the
 fat solid (modified spectral function) and the dash-dotted
 (dropping mass only) curves give the results of calculations
 \cite{Rapp} employing an in-medium modified spectral function of the vector
 mesons.} \label{CERES}
\end{minipage}

\end{center}

\end{figure}

Dileptons, i.e.\ electron-positron pairs, in the outgoing channel
are an ideal probe for in-medium properties of hadrons since they
experience no strong final state interaction. A first experiment
to look for these dileptons in heavy-ion reactions was the DLS
experiment at the BEVALAC in Berkeley \cite{DLS}. Later on, and in
a higher energy regime, the CERES experiment has received a lot of
attention for its observation of an excess of dileptons with
invariant masses below those of the lightest vector mesons
\cite{CERES}. Explanations of this excess have focused on a change
of in-medium properties of these vector mesons in dense nuclear
matter (see e.g.\ \cite{Cassingdil,RappWam}). The radiating
sources can be nicely seen in Fig.~\ref{CERES} that shows the
dilepton spectrum obtained in a low-energy run at 40 AGeV together
with the elementary sources of dilepton radiation.

The figure exhibits clearly the rather strong contributions of the
vector mesons -- both direct and through their Dalitz decay -- at
invariant masses above about 500 MeV. The strong amplification of
the dilepton rate at small invariant masses $M$ caused by the
photon propagator, which contributes $\sim 1/M^4$ to the cross
section, leads to a strong sensitivity to changes of the spectral
function at small masses. Therefore, the excess observed in the
CERES experiment can be explained by such changes as has been
shown by various authors (see e.g.\ \cite{Brat-Cass} for a review
of such calculations).

In view of the uncertainties in interpreting these results
discussed earlier we have studied the dilepton photo-production in
reactions on nuclear targets. Looking for in-medium changes in
such a reaction is not \emph{a priori} hopeless: Even in
relativistic heavy-ion reactions only about 1/2 of all dileptons
come from densities larger than $2 \rho_0$ \cite{Brat-Cass}. In
these reactions the pion-density gets quite large in the late
stages of the collision. Correspondingly many $\rho$ mesons are
formed (through $\pi + \pi \to \rho$) late in the collision, where
the baryonic matter expands and its density becomes low again.

\begin{figure}[h]

\begin{center}

\begin{minipage}[t]{7 cm}
\centerline{\epsfig{file=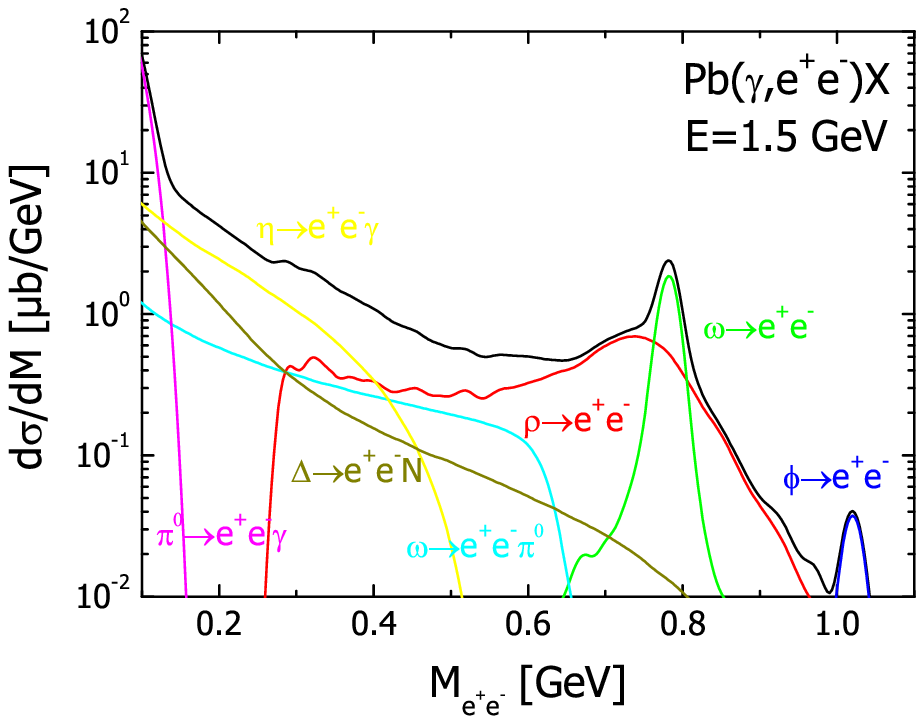,scale=1.0}}
\centerline{\epsfig{file=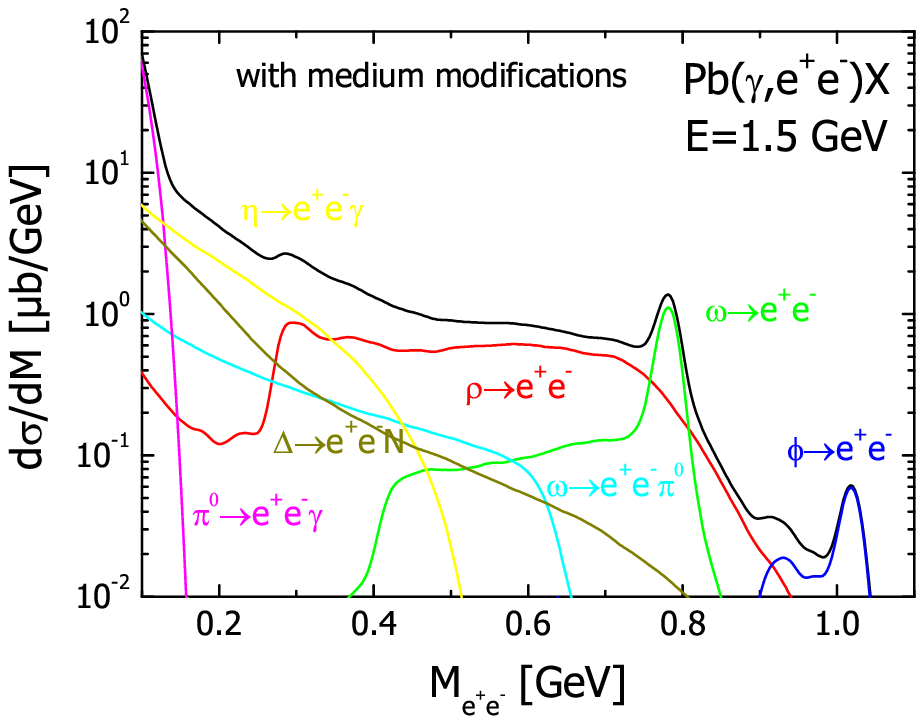,scale=1.0}}
\end{minipage}
\begin{minipage}[t]{16.5 cm}
\caption{Hadronic contributions to dilepton invariant mass spectra
for $\gamma + ^{208}Pb$ at 1.5 GeV photon energy). Indicated are
the individual contributions to the total yield; compare with
Fig.~\ref{CERES}.} \label{Fige+e-}
\end{minipage}

\end{center}

\end{figure}

\ref{Fige+e-}.
In \cite{Effephot} we have analyzed the photoproduction of
dileptons on nuclei in great detail. After removing the
Bethe-Heitler contribution the dilepton mass spectrum in a 2 GeV
photon-induced reaction looks very similar to that obtained in an
ultrarelativistic heavy-ion collision (Fig.\ \ref{CERES}). The
radiation sources are all the same in both otherwise quite
different reactions. The photon-induced reaction can thus  be used
as a baseline experiment that allows one to check crucial input
into the simulations of more complicated heavy-ion collision.

A typical result of such a calculation for the dilepton yield --
after removing the Bethe-Heitler component -- is given in Fig.\
\ref{Fige+e-}. The lower part of Fig.\ \ref{Fige+e-} shows that we
can expect observable effect of possible in-medium changes of the
vector meson spectral functions in medium on the low-mass side of
the $\omega$ peak. In \cite{Effephot} we have shown that these
effects can be drastically enhanced if proper kinematic cuts are
introduced that tend to enhance the in-medium decay of the vector
mesons. There it was shown that in the heavy nucleus $Pb$ the
$\omega$-peak completely disappears from the spectrum if in-medium
changes of width and mass are taken into account. The sensitivity
of such reactions is thus as large as that observed in
ultrarelativistic heavy-ion reactions.

An experimental verification of this prediction would be a major
step forward in our understanding of in-medium changes. The
ongoing G7 experiment at JLAB is presently analyzing such data
\cite{Weygand}. This experiment can also yield important
information on the time-like electromagnetic formfactor of the
proton and its resonances \cite{MoselHirsch95} on which little or
nothing is known.

\section{Hadron Formation}
Hadron production in reactions of high energy photons and
leptons on complex nuclei give us insight into the physics of hadron
formation. The products of a (virtual) photon-nucleon interaction
need a finite time to evolve to physical hadrons. Since the hadronization
process itself is of nonperturbative nature it is not well understood.
The formation time of a hadron may be estimated by the time that its
constituents need to travel a hadronic radius and, hence, should be
of the order of 0.5--0.8 fm/c in the rest frame of the hadron. Due to time
dilatation the formation lengths in the lab frame can become quite
large. For high energy particles they can easily
exceed the nuclear radius. Therefore, a heavy nucleus can serve as a kind of
'microdetector' which is situated directly behind the interaction
vertex and which probes the interaction of the reaction products prior to hadronization.
Hence, high energy (virtual) photonuclear reactions provide us
with essential information on the time scales of hadron formation
as well as the nature of prehadronic final state interactions.
Obviously, the latter is closely related to color transparency.

The HERMES collaboration \cite{Hermes} has studied hadron production
in deep inelastic lepton
scattering off various nuclei. Depending on the lepton beam energy
used at HERA the photon-energies are of the order of $\nu=2$--20
GeV, with rather moderate $Q^2 \approx$ 1--2 GeV$^2$. A similar
experiment is currently performed by the CLAS collaboration at
Jefferson Lab \cite{Brooks} using a 5 GeV electron beam. After the
planned energy upgrade they will reach photon energies of the order
of $\nu=2$--9 GeV and similar virtualities as in the HERMES experiment.

The observed multiplicity spectra have led to a numerous different
interpretations. They reach from a purely partonic energy loss through
induced gluon radiation of the struck quark propagating through
the nucleus \cite{ArleoWang} to a possible rescaling of the quark
fragmentation function at finite nuclear density combined with hadron
absorption in the nuclear environment \cite{Accardi}. Note that the
detailed understanding of hadron attenuation in cold nuclear matter can
serve as a useful basis for the interpretation of jet quenching
observed in ultra-relativistic heavy ion collisions at RHIC.

In our approach \cite{Falterneu,Falteralt} we assume that the virtual photon
interacts with the bound nucleon either directly or via one of its
hadronic fluctuations. The hadronic components of the photon are
thereby shadowed according to the method described in Sec.~\ref{sec:CoupledChannel}
which allows for a clean-cut separation of the
coherent initial state interactions of the photon and the
incoherent final state interactions of the reaction products. The
photon-nucleon interaction leads to the excitation of one or more
strings which fragment into color-neutral prehadrons due to the
creation of quark-antiquark pairs from the vacuum. As discussed by
Kopeliovich et al. \cite{Kopeliovich} the
production time of these prehadrons is very short. For simplicity
we set the production time to zero in our numerical realization.
These prehadrons are then propagated using our coupled-channel
transport theory.

After a formation time which we assume to be a constant $\tau_f$
in the restframe of the hadron the hadronic wave function has
built up and the reaction products behave like usual hadrons. The
prehadronic cross sections $\sigma^*$ during the formation time
are determined by a simple constituent quark model
\begin{eqnarray}
\label{eq:prehadrons} \sigma^*_\mathrm{prebaryon}&=
&\frac{n_\mathrm{org}}{3}\sigma_\mathrm{baryon} , \nonumber\\
    \sigma^*_\mathrm{premeson}&=&\frac{n_\mathrm{org}}{2}\sigma_\mathrm{meson}
    ,
\end{eqnarray}
where $n_\mathrm{org}$ denotes the number of (anti-)quarks in the
prehadron stemming from the beam or target. As a consequence the
prehadrons that solely contain (anti-)quarks produced from the
vacuum in the string fragmentation do not interact during
$\tau_f$. Using this recipe the total effective cross section of
the final state rises like in the approach of Ref. \cite{Ciofi}
each time when a new hadron has formed.
\begin{figure}[h]
\vspace*{-0.0cm}
 \centering{\includegraphics[width=15.0cm]{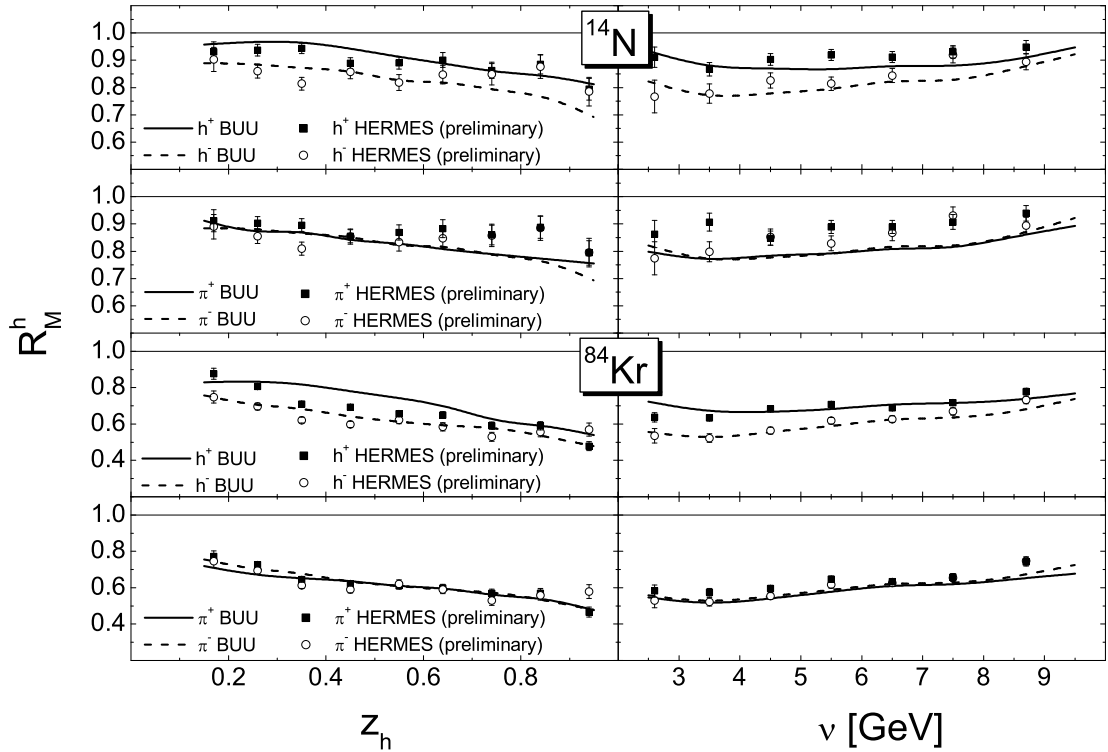}}
\vspace*{-0.5cm} \caption{Calculated multiplicity ratio of
positively and negatively charged hadrons and pions for a $^{14}$N
and $^{84}$Kr target when using a 12 GeV positron beam at HERMES.
For the calculation we use the formation time $\tau_f=0.5$ fm/$c$
and the constituent-quark concept (\ref{eq:prehadrons}) for the
prehadronic cross sections. The data are taken from
Ref.~\cite{Nez04}.} \label{fig:HERMES12}
\end{figure}

Due to our coupled channel-treatment of the final state
interactions the (pre)hadrons might not only be absorbed in the
nuclear medium but can produce new particles in an interaction,
thereby shifting strength from the high to the low energy part of
the hadron spectrum. In addition our event-by-event simulation
allows us to account for all sorts of kinematic cuts and the
acceptance of the detector. In Ref.~\cite{Falterneu} we have
demonstrated that our calculations are in excellent agreement with
the experimental HERMES data \cite{Hermes} taken on various
nuclear targets at beam energy $E_\mathrm{beam}=27.6$ GeV if one
assumes a formation time $\tau_f=0.5$ fm/c for all hadron species.

In Fig.~\ref{fig:HERMES12} we show that our approach is also
capable to describe the observed multiplicities of charged hadrons
and pions at half the possible beam energy in the HERMES
experiment, i.e.~at $E_\mathrm{beam}=$12 GeV. This makes us
confident that our model can also be applied for the electron beam
energies that will be used at Jefferson Lab.

\begin{figure}[h]
\vspace*{-0.0cm}
 \centering{\includegraphics[width=10.0cm]{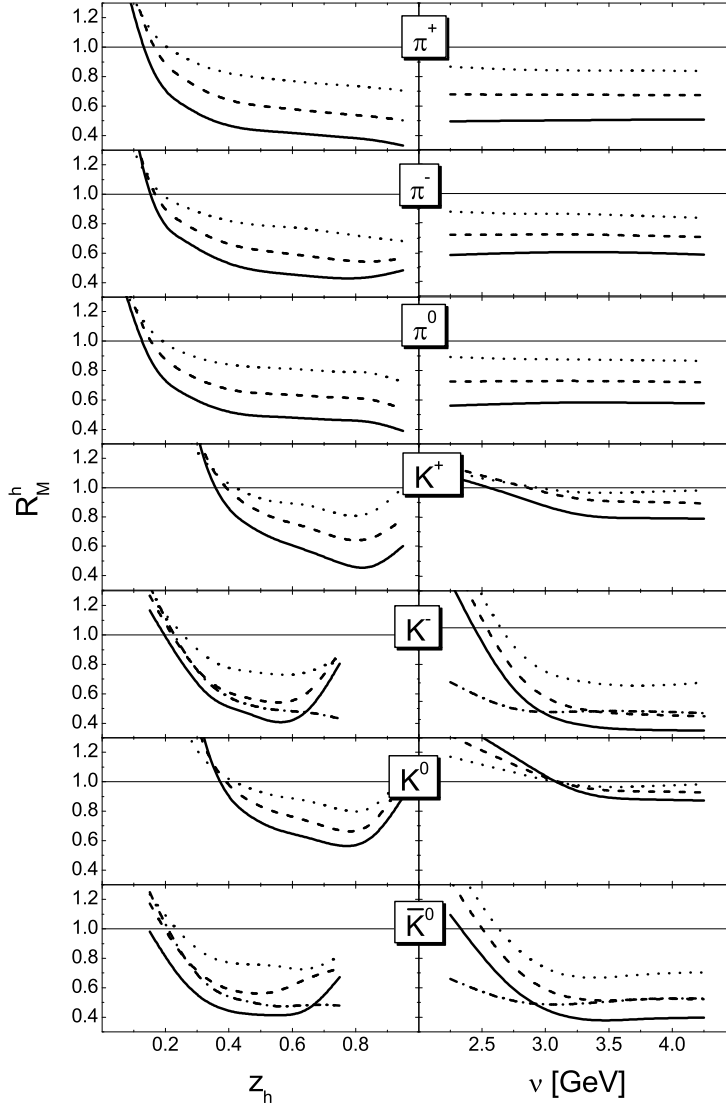}}
\vspace*{-0.5cm}

\caption{Calculated multiplicity ratio of identified $\pi^\pm$,
$\pi^0$, $K^\pm$, $K^0$ and $\bar{K}^0$ for $^{12}$C (dotted
lines), $^{56}$Fe (dashed lines) and $^{208}$Pb nuclei (solid
lines). The simulation has been done for a 5 GeV electron beam and
the CLAS detector. The dash-dotted line represents a calculation
for $^{56}$Fe without Fermi motion. In all calculations we use the
formation time $\tau_f=0.5$ fm/$c$ and the constituent-quark
concept (\ref{eq:prehadrons}) for the prehadronic cross sections
(from \cite{FalterPhD}).} \label{fig:JLab}
\end{figure}

In Fig.~\ref{fig:JLab} we present our predictions for the
multiplicity ratios of identified hadrons at 5 GeV electron beam
energy. Besides the considerably lower beam energy used at
Jefferson Lab the major difference to the HERMES experiment is the
much larger geometrical acceptance of the CLAS detector. The
latter leads to an increased detection of low energy secondary
particles that are produced in the final state interactions and
that lead to a strong increase of the multiplicity ratio at low
fractional hadron energies $z_h=E_h/\nu$. In addition, the
relatively small average photon energy leads to a visible effect
of Fermi motion on the multiplicity ratio of more masive
particles. Since the virtual photon cannot produce antikaons
without an additional strange meson, e.g.~$\gamma^*N\to
K\bar{K}N$, the maximum fractional energy $z_h$ is limited to 0.9
for antikaons. Because of the energy distribution in the three
body final state and the finite virtuality of the photon -- set by
the kinematic cut $Q^2>1$ GeV$^2$ -- the maximum fractional energy
of antikaons is further reduced. As a result, the $z_h$ spectra
for $K^-$ and $\bar{K}^0$ in Fig.~\ref{fig:JLab} do not exceed
$z_h\approx0.8$. For the same reason the production of antikaons
with $z_h>0.2$ is reduced at the lower end of the photon spectrum.
The Fermi motion in the nucleus enhances the yield of antikaons in
these two extreme kinematic regions as can be seen by comparison
with the dash-dotted line in Fig.~\ref{fig:JLab} which represents
the result of a calculation for $^{54}$Fe where Fermi motion has
been neglected. Certainly, the kaons can be produced in a two body
final state (e.g.~$\gamma^*N\to K\Lambda$), however, the
accompanying hyperon has a relatively large mass. Therefore,
similar, although less pronounced, effects show also up for the
kaons. Beside the effects of Fermi motion the multiplicity ratios
of kaons and antikaons show the same features as for higher
energies.

\section{Conclusions}\label{concl}
In this lecture note we have shown that photonuclear reactions on
nuclei can give observable consequences of in-medium changes of
hadrons that are as big as those expected in heavy-ion collisions
which reach much higher energies, but proceed farther away from
equilibrium. Information from photonuclear reactions is important
and relevant for an understanding of high density -- high
temperature phenomena in ultrarelativistic heavy-ion collisions.
Special emphasis was put in this article not so much on the
theoretical calculations of hadronic in-medium properties under
simplified conditions, but more on the final, observable effects
of any such properties. We have discussed that for reliable
predictions of observables one has to take the final state
interactions with all their complications in a coupled channel
calculation into account; simple Glauber-type descriptions are not
sufficient. We have, for example, shown that in photonuclear
reactions in the 1 - 2 GeV range the expected sensitivity of
dilepton spectra to changes of the $\rho$- and $\omega$ meson
properties in medium is as large as that in ultrarelativistic
heavy-ion collisions and that exactly the same sources contribute
to the dilepton yield in both experiments. We have also
illustrated that the analysis of hadron production spectra in
high-energy electroproduction experiments at HERMES gives
information about the interaction of forming hadrons with the
surrounding hadronic matter. This is important for any analysis
that tries to obtain signals for a QGP by analysing high-energy
jet formation in ultrarelativistic heavy-ion reactions.

\section*{Acknowledgement}
This work has been supported by the Deutsche
Forschungsgemeinschaft, the BMBF and GSI Darmstadt.

\end{document}